\pdfoutput=1

\documentclass[11pt]{article}

\usepackage[longnamesfirst]{natbib} 
\usepackage{amsmath} 
\usepackage{amsfonts} 
\usepackage{amsthm}
\usepackage{amssymb}
\usepackage{graphicx} 
\usepackage{xcolor} 
\usepackage{xspace}
\usepackage{ifmtarg} 
\usepackage{rotating}
\usepackage[normalem]{ulem}
\usepackage{slashbox} 

\RequirePackage[colorlinks,citecolor=black,urlcolor=black]{hyperref} 
\hypersetup{pdfauthor={J.-P. Baudry},pdftitle={Enhancing the selection of a model-based clustering with external qualitative variables}}

\newcommand{\bu}{\mathbf{u}}
\newcommand{\buone}{\mathbf{u^1}}
\newcommand{\buj}{\mathbf{u^j}}
\newcommand{\bur}{\mathbf{u^r}}

\newcommand{\bz}{\mathbf{z}}

\newcommand{\by}{\mathbf{y}}

\definecolor{remtxt}{gray}{0.4}
\definecolor{rem2}{rgb}{1,0,0}
\definecolor{propcol}{gray}{0.4}

\newcommand{\prop}[2][] 
   {\textcolor{propcol}{#2}} 
\newcommand{\ok}[2][]{#2} 
\newcommand{\fig}[1]{Figure~\ref{#1}}
\newcommand{\tab}[1]{Table~\ref{#1}}
\newcommand{\sect}[1]{Section~\ref{#1}}

\newcommand{\dens}{f}

\DeclareMathOperator*{\argmax}{argmax}

\title{Enhancing the selection of a model-based clustering with external categorical variables}

\author{		Jean-Patrick Baudry \footnote{LSTA, Universit\'e Pierre et Marie Curie -- Paris VI} 
		\and
		 	Margarida Cardoso \footnote{BRU-UNIDE, ISCTE-IUL}
		\and 
			Gilles Celeux \footnote{INRIA Saclay-\^Ile-de-France}
		\and 
			Maria Jos\'e Amorim \footnote{ISEL, ISCTE-Lisbon University Institute}
		\and 
			Ana Sousa Ferreira \footnote{ProjAVI (MEC), BRU-UNIDE \& CEAUL}
}

\begin{document}

\maketitle

\begin{abstract}

In cluster analysis, it can be useful to interpret the partition built from the data in the light of external categorical variables which were not directly involved to cluster the data. An approach is proposed in the model-based clustering context to select a model and a number of clusters which both fit the data well and take advantage of the potential illustrative ability of the external variables. This approach makes use of the integrated joint likelihood of the data and the partitions at hand, namely the model-based partition and the partitions associated to the external variables. It is noteworthy that each mixture model is fitted by the maximum likelihood methodology to the data, excluding the external variables which are used to select a relevant mixture model only. Numerical experiments illustrate the promising behaviour of the derived criterion.

\end{abstract}

\paragraph{Keywords} Mixture Models; Model-Based Clustering; Number of Clusters; Penalised Criteria; Categorical Variables; BIC; ICL

\clearpage

\section{Introduction}

In model selection, assuming that the data arose from one of the models in
competition is often somewhat unrealistic and could be misleading. However this assumption is implicitly
made when using standard model selection criteria such as AIC or BIC.
This ``true model'' assumption could lead to overestimating the model complexity in practical situations.
On the other hand, a common feature of standard penalized likelihood criteria such as AIC and BIC is that they do not take into
account the modelling purpose. Our opinion is that it is worthwhile taking it into account to select a
model, which leads to more flexible criteria favoring useful and
parsimonious models. This point of view could be exploited in many statistical learning situations.

Whereas cluster analysis is an exploratory data analysis tool, any available information on the objects to be clustered, available in addition to the {\it clustering} variables, could be very useful to get a meaningful interpretation of the clusters. Here we address the case where this additional information is provided by external categorical {\it illustrative} variables. The purpose of this paper is to introduce a model selection criterion in the model-based clustering context that takes advantage of these illustrative variables. This criterion aims to select a classification of the data which achieves a good compromise: it is expected to provide a parsimonious and sensible clustering with a relevant interpretation with respect to the illustrative categorical variables. It is important to stress that we do not want the external variables to affect the classifications derived from the clustering variables: they are merely used to highlight some of them.

The paper is organised as follows. In \sect{sic.mbc}, the framework of model-based clustering is described. Our new penalised likelihood criterion is presented in \sect{sic.sic}.
Numerical experiments on simulated and real data sets are presented in \sect{sic.sim} to illustrate the behavior of this criterion and highlight its possible interest. A short
discussion section concludes the paper.

\section{Model-based clustering} \label{sic.mbc}

Model-based clustering consists of modelling the data to be classified by a mixture distribution and of associating a class with each of the mixture components. Embedding cluster analysis in this precise framework is useful
in many aspects. In particular, it allows to choose the number $K$ of classes (i.e. the number of mixture components) rigorously.

\subsection{Finite mixture models}

Please refer to \citet{MclPee00} for a comprehensive introduction to finite mixture models.

The data to be clustered ${\by} = (\by_1,\dots,\by_n) $, with $\by_i \in \mathbf R^d$, are modelled as observations of iid random variables with a mixture distribution: 
\begin{equation*}
\dens({\mathbf y} \mid \theta)=\prod_{i=1}^{n} \dens ({\mathbf y}_i \mid \theta)\ \text{with}\ \dens({\mathbf y}_i \mid \theta)= \sum_{k=1}^{K} p_{k}\phi({\mathbf y}_i \mid {\mathbf a}_{k})
\end{equation*}
where the $p_{k}$'s are the mixing proportions and $\phi(\cdot \mid {\mathbf a}_{k})$ denotes the components probability density function
(typically the $d$-dimensional Gaussian density)
with parameter ${\bf a}_{k}$, and $\theta=(p_1,\ldots,p_{K-1},{\mathbf a}_1,\ldots,{\mathbf a}_K)$. A mixture model can be regarded as a latent structure model involving unknown label data ${\mathbf z}=({\mathbf z}_1,\ldots,{\mathbf z}_n)$ which
are binary vectors with $z_{ik}=1$ if and only if ${\mathbf y}_i$ arises from component $k$.
Those indicator
vectors define a partition $P=(P_1,\ldots,P_K)$ of the data ${\mathbf y}$ with $P_k=\{{\mathbf y}_i
\mid z_{ik}=1\}$. However these indicator vectors are not observed in a clustering problem:
the model is usually fitted through maximum likelihood estimation and an estimated partition is deduced from it by the MAP rule recalled in \eqref{MAP}. The parameter estimator, denoted from now on by $\hat\theta$, is generally derived from the EM algorithm \citep{DemLaiRub77, MclKri97}. 

Remark that, for a given number of components $K$ and a parameter $\theta_K$, the class of each observation $\by_i$ is assigned according to the MAP rule defined above.

There are usually several models to choose among (typically, when the number of components is unknown). Note that a mixture model $m$ is characterized not only by the number of components $K$, but also by assumptions on the proportions and the component variance matrices \citep[see][]{CelGov95}. The corresponding parameter space is denoted by $\Theta_m$. From a density estimation perspective, a classical way for choosing a mixture model is to select the model maximising the integrated likelihood,
\[
\dens({\mathbf y} \mid m)=\int_{\Theta_m}\dens({\mathbf y} \mid m, \theta_m)
\pi(\theta_m) d\theta_m,
\]
$\pi(\theta_m)$ being a weakly informative prior distribution
on $\theta_m$.
For $n$ large enough, it can be approximated with the BIC criterion \citep{Sch78}
\[
\log \dens ({\mathbf y} \mid m) \approx \log \dens({\mathbf y} \mid  m, \hat\theta_m)-\frac{\nu_{m}}{2} \log n,
\]
with $\nu_m$ the number of free parameters in the mixture model $m$. Numerical experiments \citep[see for instance][]{RoeWas97} and theoretical results \citep[see][]{Ker00} show
that BIC works well to select the true number of components when the data actually arises from one of the mixture models in competition.

\subsection{Choosing $K$ from the clustering view point}

In the model-based clustering context, an alternative to the BIC criterion
is the ICL criterion \citep{BieCelGov00} which aims at maximising
the integrated likelihood of the complete data $({\mathbf y}, {\mathbf
z})$
\[
\dens ({\mathbf y}, {\mathbf z} \mid m)=\int_{\Theta_{m}} \dens ({\mathbf y},
{\mathbf z} \mid m, \theta_m)
\pi(\theta_m) d\theta_m.
\]
It can be approximated with a BIC-like approximation:
\[
\log \dens({\mathbf y},{\mathbf z} \mid m)\approx
\log \dens({\mathbf y},{\mathbf z} \mid m, \hat\theta_m^{*}) - \frac{\nu_{m}}{2} \log n
\]
\[
\hat \theta_m^{*}=\arg \max_{\theta_m} \dens({\mathbf y},{\mathbf z} \mid m, \theta_m).
\]
But ${\mathbf z}$ and $\hat\theta_m^{*}$ are unknown. Arguing that $\hat \theta_m \approx \hat \theta_m^{*}$ if the mixture components are well separated for $n$ large enough, \citet{BieCelGov00} replace $\hat \theta_m^{*}$ by $\hat \theta_m$ and
the missing data ${\mathbf z}$ with $\hat {\mathbf  z}=\mbox{MAP}(\hat \theta_m)$
defined by
\begin{equation}\label{MAP}
\hat z_{ik}= \left\{ \begin{array}{ll} 1 & \mbox{if } \mbox{argmax}_{\ell}\,  \tau_i^\ell(\hat \theta_m)=k\\ 0 & \mbox{otherwise,} \end{array} \right.
\end{equation}
$\tau_i^k(\theta_m)$ denoting the conditional probability that ${\mathbf y}_{i}$ arises from the $k$th mixture component under $\theta_m$ $(1\leq i \leq n$ and $1\leq k \leq K)$:
\begin{equation}
\tau_i^k(\theta_m)=\frac{p_k \phi({\mathbf y}_i \mid {\mathbf a}_k)}{\sum_{\ell=1}^K p_\ell \phi({\mathbf y}_i \mid {\mathbf a}_\ell)}.
\end{equation}
Finally the ICL criterion is
\begin{equation}\label{def_icl}
\mbox{ICL}(m)= \log \dens({\mathbf y}, \hat {\mathbf z} \mid m,\hat \theta_m)
- \frac{\nu_{m}}{2} \log n.
\end{equation}
Roughly speaking ICL is the criterion BIC decreased by the
estimated mean entropy
\[
E(m)=-\sum_{k=1}^{K}\sum_{i=1}^{n}\tau_i^k(\hat \theta_m) \log \tau_i^k(\hat \theta_m) \geq 0.
\]
This is apparent if the estimated labels $\hat \bz$ are replaced in the definition \eqref{def_icl} by their respective conditional expectation $\tau_i^k(\hat \theta_m)$, since $\log f(\by,\bz | m, \theta_m) = \log f(\by | m, \theta_m) +  \sum_{i=1}^n \sum_{k=1}^K z_{ik} \log \tau_i^k (\theta_m)$.

Because of this additional entropy term, ICL favors models which lead to partitioning the
data with the greatest evidence. The derivation and approximations leading to ICL are questioned in \citet[][Chapter 4]{Bau09}.
However, in practice, ICL appears to provide a stable and
reliable estimate of the number of mixture components for real data sets and also for simulated data sets \ok[arising from mixtures
with well separated components]{from the clustering view point}. ICL, which is not aiming at discovering the true number of mixture
components, can underestimate the number of components for simulated data arising from
mixture\ok{s} with poorly separated components \citep{BieCelGov00}. \ok{It concentrates on selecting a relevant number of classes.}

\section{A particular clustering selection criterion}\label{sic.sic}


Now, suppose that, beside $\by$, a known classification $\bu$ (e.g. associated to an extra categorical variable) is available. We still want to build a classification $\bz$ based on $\by$, which is supposed to carry some more information than $\bu$. But relating the classifications $\bz$ and $\bu$ could be of interest to get a suggestive and simple interpretation of $\bz$. Therefore, we propose to build the classification $\bz$ in each model, based on $\by$ only, but to involve $\bu$ in the model selection step. Hopefully $\bu$ might highlight some of the solutions among which $\by$ would not enable to decide clearly. This might help to select a model providing a good compromise between the mixture model fit to the data and its ability to lead to a classification of the observations well related to the external classification $\bu$. To derive our heuristics, we suppose that $\by$ and $\bu$ are conditionally independent knowing $\bz$, which means that all the relevant information in $\bu$ and $\by$ can be caught by $\bz$. This is for example true in the very particular case where $\bu$ can be written as a function of $\bz$: $\bu$ is a reduction of the information included in $\bz$, and we hope to be able to retrieve more information from $\bu$ using the (conditionally independent) information brought by $\by$. 

Here is our heuristics. It is based on an intent to find the mixture model maximizing the integrated completed likelihood
\begin{equation}\label{integratedp}
\dens(\by,\bu,\bz \mid  m) = \int \dens(\by,\bu,\bz \mid  m, \theta_m) \pi(\theta_m) d\theta_m.
\end{equation}

Assuming that $\by$ and $\bu$ are conditionally independent knowing $\bz$, which should hold at least for models with enough components, it can be written for any $\theta_m\in m$:
\begin{equation}\label{eq.der1}
\dens(\by,\bu,\bz | m, \theta_m) = \dens(\by, \bz | m, \theta_m) \underbrace{\dens(\bu | \by, \bz, m, \theta_m)}_{\dens(\bu | \bz,m,\theta_m)}.
\end{equation}
But neither $\theta_m$ nor $m$ carry any information on the model for $\bu | \bz$ and then $\dens(\bu | \bz,m, \theta_m) = \dens(\bu|\bz)$. Let us denote $(n_{k\ell})_{\substack{1\leq \ell\leq U, 1\leq k\leq K}}$ the contingency table relating the categorical variables $\bu$ and $\bz$: for any $k\in\{1,\ldots,K\}$ and $\ell\in \{1,\ldots,U\}$, $U$ being the number of levels of the variable $\bu$,
\begin{equation*}
n_{k\ell} = \mbox{card} \bigl\{ i | z_{ik}=1 \mbox{ and } u_i=\ell \bigr\}.
\end{equation*}
Moreover, let us denote $n_{k.}=\sum_{\ell=1}^{U} n_{k\ell}$. Denoting
\begin{equation*}
\mathcal L = \bigl\{ (q_{k\ell})_{\substack{1\leq \ell\leq U \\ 1\leq k\leq K }} \in (0,1)^{K \times U} | \forall k\in\{1,\dots,K\}, \sum_{\ell = 1}^{U} q_{k\ell} = 1 \bigr\},
\end{equation*}
we have 
\begin{equation*}
\argmax_{(q_{k\ell})\in\mathcal L} \sum_{i = 1}^n \log q_{z_i u_i} = \Bigl(\frac{n_{k\ell}}{n_{k .}} \Bigr)_{\substack{1\leq \ell\leq U \\ 1\leq k\leq K }}.
\end{equation*}
Thus, we get 
\begin{align*}
\log \dens(\bu \mid \bz)  & = \sum_{i=1}^n \log\frac{n_{z_i u_i}}{n_{z_i .}} \\
&= \sum_{\ell=1}^{U}\sum_{k=1}^K n_{k\ell}\log\frac{n_{k\ell}}{n_{k.}}
\end{align*}
and, from \eqref{integratedp} and \eqref{eq.der1},
\begin{equation*}\label{eq_SICL}
\log \dens(\by,\bu,\bz \mid  m) = \sum_{\ell=1}^{U}\sum_{k=1}^K n_{k\ell}\log\frac{n_{k\ell}}{n_{k.}} + \log \int \dens(\by,\bz \mid  m, \theta_m) \pi(\theta_m) d\theta_m.
\end{equation*}
Now, $\log \int \dens(\by,\bz \mid  m, \theta_m) \pi(\theta_m) d\theta_m$ can be approximated by ICL as in \eqref{def_icl}. 
Thus
\begin{equation*}
\log \dens(\by,\bu,\bz \mid  m ) \approx \log \dens(\by,\bz\mid m,\hat\theta_m) + \sum_{\ell=1}^{U}\sum_{k=1}^K n_{k\ell}\log\frac{n_{k\ell}}{n_{k.}} - \frac{\nu_m}{2}\log n.
\end{equation*}

Finally, this leads to the Supervised Integrated Completed Likelihood (SICL) criterion
$$
SICL(m)=\mbox{ICL}(m)+\sum_{\ell=1}^{U} \sum_{k=1}^{K} n_{k \ell} \log \frac{n_{k\ell}}{n_{k \cdot}}.
$$
The last additional term $\sum_{\ell=1}^{U} \sum_{k=1}^{K} n_{k \ell} \log \frac{n_{k\ell}}{n_{k \cdot}}$ quantifies the strength of the link between the categorical variables $\bu$ and $\bz$. This might be helpful eventually for the interpretation of the classification $\bz$.

%

\paragraph{Taking several external variables into account}
The same kind of derivation enables to derive a criterion that takes into account several external variables $\buone, \dots, \bur$. Suppose that $\by,\buone,\dots, \bur$ are conditionally independent knowing $\bz$. Then \eqref{eq.der1} becomes
\begin{equation}
\begin{split}
\dens(\by,\buone,\dots, \bur, \bz \mid m,\hat\theta^*_m)&= \dens(\by, \bz \mid m,\hat\theta^*_m)\\
&\qquad\quad \times \underbrace{\dens(\buone \mid \by, \bz,m,\hat\theta^*_m)}_{\dens(\buone \mid \bz,m,\hat\theta^*_m)}\\
&\qquad\quad\quad \times \dots\\
&\qquad\quad\quad\quad \times \underbrace{\dens(\bur \mid \by, \bz,m,\hat\theta^*_m)}_{\dens(\bur \mid \bz,m,\hat\theta^*_m)},
\end{split}
\end{equation}
with $\hat \theta_m^{*}=\arg \max_{\theta_m} \dens({\mathbf y},\buone,\dots,\bur,{\mathbf z} \mid m,\theta_m)$. As before, we assume that $\hat \theta_m \approx \hat \theta_m^*$ and apply the BIC-like approximation. Finally,
\begin{align*}
\log \dens(\by,\buone,\dots,\bur,\bz \mid  m ) &\approx \log \dens(\by,\bz\mid m,\hat\theta_m) \\
&\qquad + \log \dens(\buone \mid \bz,m,\hat \theta_m) \\
&\qquad\quad + \dots \\
&\qquad\quad\quad + \log \dens(\bur \mid \bz,m,\hat \theta_m) \\
&\qquad\quad\quad\quad -\frac{\nu_m}{2}\log n,
\end{align*}
and as before, $\log \dens(\buj \mid  \bz) = \log \dens(\buj \mid  \bz, m, \hat \theta_m)$ is derived from the contingency table $(n_{k\ell}^j)$ relating the categorical variables $\buj$ and $\bz$: for any $k\in\{1,\ldots,K\}$ and $\ell\in \{1,\ldots,U^j\}$, $U^j$ being the number of levels of the variable $\buj$,
\begin{equation*}
n_{k\ell}^j = \mbox{card} \bigl\{ i | z_{ik}=1 \mbox{ and } u_i^j=\ell \bigr\}.
\end{equation*}
Finally, with $n_{k.} = \sum_{\ell=1}^{U^j} n_{k\ell}^j$, which does not depend on $j$, we get the ``multiple'' external variables criterion:
\begin{equation*}
\mbox{SICL}(m)=\mbox{ICL}(m)+\sum_{j=1}^r \sum_{\ell=1}^{U^j} \sum_{k=1}^{K} n_{k \ell}^j \log \frac{n_{k\ell}^j}{n_{k \cdot}}.
\end{equation*}

\section{Numerical experiments}\label{sic.sim}

All the numerical experiments rely on the {\it Rmixmod} package for R \citep{LebIovLanBieCelGov12}. We first present an illustration of the behaviour of SICL on the Iris data set. The behaviour of SICL is then analysed in various situations by numerical experiments on simulated data. Finally an application on a real data set is presented.

\subsection{Illustrative numerical experiment}

\begin{figure}[htbp]
	\begin{minipage}[c]{0.48\linewidth}
		\begin{center}
			 \includegraphics[trim=20cm 0.7cm 0 5cm, clip, width = 1.5\linewidth]{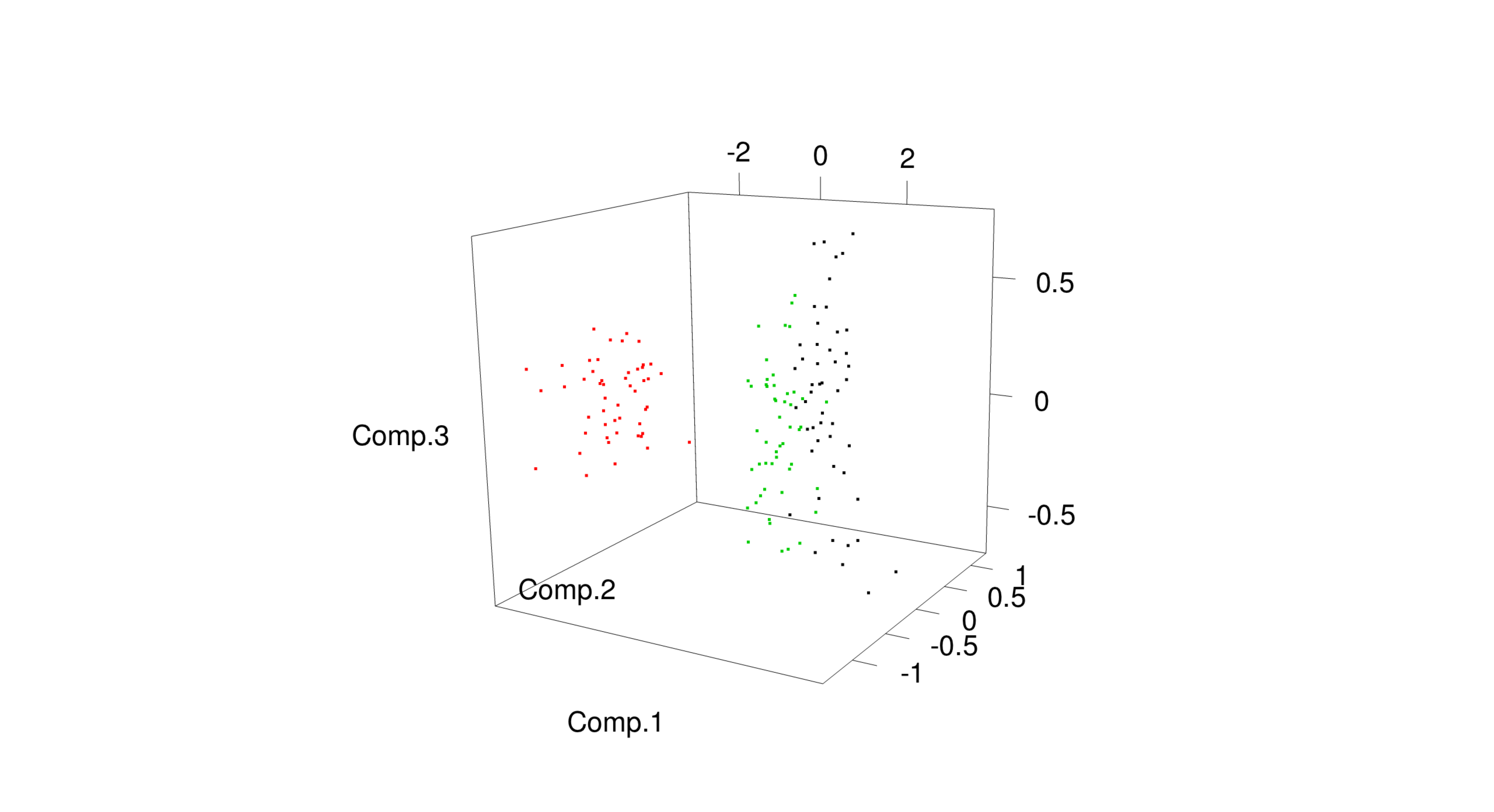}
		\end{center}
	\end{minipage}
	\begin{minipage}[c]{0.48\linewidth} 
		\vspace{-2em}
		\begin{center}
			\includegraphics[trim=0cm 0.7cm 0 0, clip]{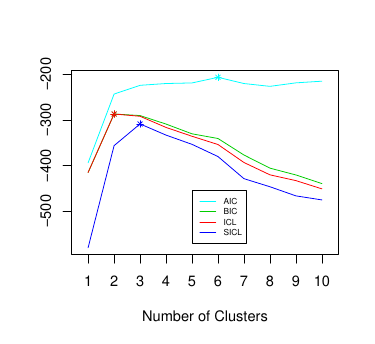}
		\end{center}
	\end{minipage}
	\caption{First three principal components for the Iris data set (left) and corresponding information criteria versus the number of clusters (right)}\label{Iris}
\end{figure}

The first example is an application to the Iris data set \citep{Fis36} which consists of $150$ observations of four measurements ($\by$)
for three species of Iris ($\bu$). These data are depicted in \fig{Iris}. The most general Gaussian mixture model was considered. The variations of criteria BIC, ICL and SICL in function of $K$
are provided in \fig{Iris}. While BIC and ICL choose two classes, SICL selects the three-component mixture solution which is closely related to the species of Iris, as attested by the contingency table between the two partitions (\tab{IrisTab}).

\begin{table}[htbp]
	\begin{center}
		\caption{\label{IrisTab} Iris data. Contingency table between the ``species'' variable and the classes derived from the three-component mixture.}
		\begin{tabular}{| l| c| c| c| }
			\hline
			\backslashbox{Species}{k} & 1 & 2 & 3 \\ 
			\hline
			Setosa & 0 & 50 & 0 \\
			Versicolor & 45 & 0 & 5 \\
			Virginica & 0 & 0 & 50 \\
			\hline
		\end{tabular}
	\end{center}
\end{table}

\subsection{Simulated numerical experiments}\label{sic.sim.sim}

\begin{figure}[htbp]
	\begin{minipage}[c]{0.48\linewidth}
		\vspace{-2em}
		\begin{center}
			 \includegraphics[trim=0.65cm 0.7cm 0 0, clip]{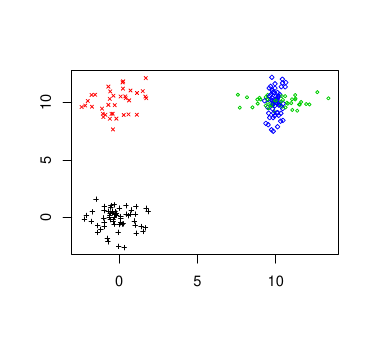}
		\end{center}
	\end{minipage}
	\begin{minipage}[c]{0.48\linewidth} 
		\vspace{-2em}
		\begin{center}
			\includegraphics[trim=0cm 0.7cm 0 0, clip]{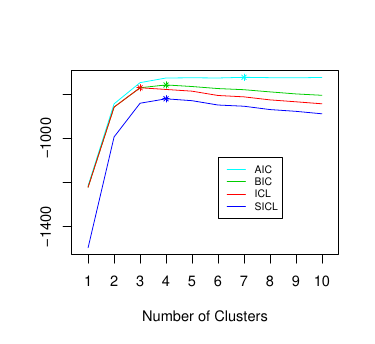}
		\end{center}
	\end{minipage}
	\caption{``Cross'' data set external classification (left) and corresponding information criteria versus the number of clusters (right)}\label{Cross}
\end{figure}

For the second experiment, we simulated $200$ observations from a Gaussian mixture in $\mathbf R^2$ depicted in \fig{Cross} and the variable $\bu$ corresponds exactly to the mixture component from which each observation arises. Diagonal mixture models (i.e. with diagonal variance matrices) are fitted.
The variations of the criteria BIC, ICL and SICL in function of $K$ are provided in \fig{Cross}. We repeated this experiment with $100$ different simulated data sets. BIC almost always recovers the four Gaussian components, while ICL almost always selects three because of the two very overlapping ones (the ``cross''). Since the solution obtained through MLE with the four-component mixture model yields classes nicely related to the considered $\bu$ classes, SICL favors the four-component solution more than ICL does. But since it also takes the overlapping into account, it still selects the three-component model $18$ times out of $100$, and selects the four-component model in almost all the remaining cases ($80$ out of $100$). SICL allows to decide between the solutions among which BIC and ICL hesitate.


\begin{figure}[htbp]
	\begin{minipage}[c]{0.48\linewidth}
		\vspace{-2em}
		\begin{center}
			 \includegraphics[trim=0.65cm 0.7cm 0 0, clip]{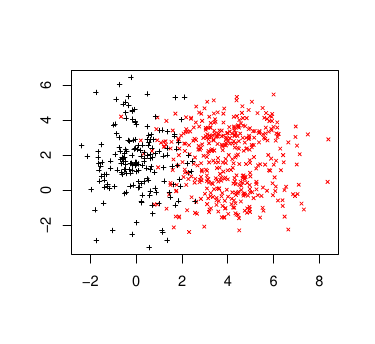}
		\end{center}
	\end{minipage}
	\begin{minipage}[c]{0.48\linewidth} 
		\vspace{-2em}
		\begin{center}
			\includegraphics[trim=0cm 0.7cm 0 0, clip]{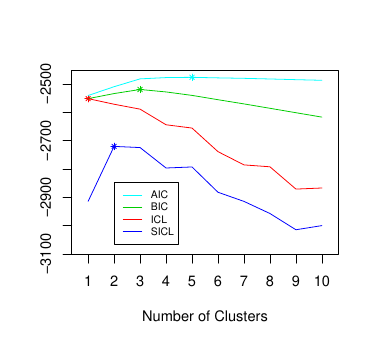}
		\end{center}
	\end{minipage}
	\caption{Third experiment data set external classification (left) and corresponding information criteria versus the number of clusters (right)}\label{Gilles}
\end{figure}

The third experiment illustrates a situation where SICL gives a relevant solution different from the solutions selected with BIC and ICL. We simulated $200$ observations of a diagonal three-component Gaussian mixture depicted in Figure~\ref{Gilles} where the classes of $\bu$ are in red and in black. (The red class is composed of two ``horizontal'' Gaussian components while the black one is composed of a single ``vertical'' Gaussian component...) The most general Gaussian mixture model was considered. From Table~\ref{GillesTab} BIC almost always recovers the three components of the simulated mixture, ICL mostly selects one cluster, and SICL chooses two classes well related to $\bu$. 

\begin{table}[htbp]
	\begin{center}
		\caption{\label{GillesTab} Number of components selected by each criterion for the third experiment data}
		\begin{tabular}{| l| c| c| c| c| c| c| c| c| c| c|}
			\hline
			K & 1 & 2 & 3 & 4 & 5 & 6 & 7 & 8 & 9 & 10 \\ 
			\hline
			AIC &0 &0 &24 &26 &20 &8 &6 &4 &5 & 7 \\
			BIC & 0 &4 &96 &0 &0 &0 &0 &0 &0 & 0 \\
			ICL & 53 &44 &3 &0 &0 &0 &0 &0 &0 & 0 \\
			SICL & 0 &86 &14 &0 &0 &0 &0 &0 &0 & 0 \\
			\hline
		\end{tabular}
	\end{center}
\end{table}

In the next two experiments, we analyse the behaviour of SICL in situations where  $\bu$ cannot be related with the mixture distributions at hand. 

\begin{figure}[htbp]
	\begin{minipage}[c]{0.48\linewidth}
		\vspace{-2em}
		\begin{center}
			 \includegraphics[trim=0.65cm 0.7cm 0 0, clip]{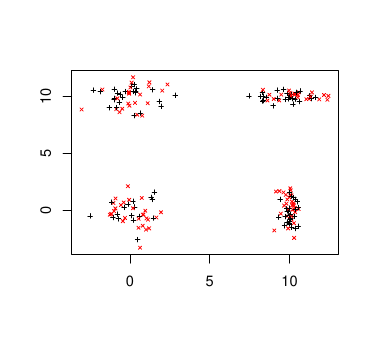}
		\end{center}
	\end{minipage}
	\begin{minipage}[c]{0.48\linewidth} 
		\vspace{-2em}
		\begin{center}
			\includegraphics[trim=0cm 0.7cm 0 0, clip]{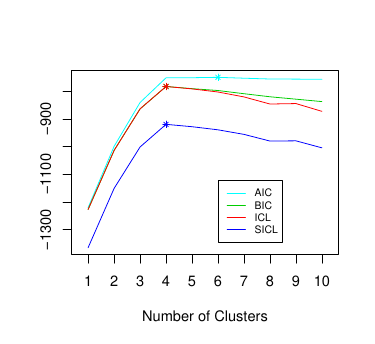}
		\end{center}
	\end{minipage}
	\caption{``Random labels'' data set external classification (left) and corresponding information criteria versus the number of clusters (right)}\label{Random}
\end{figure}

At first we consider a situation where $\bu$ is a two-class partition which has
no link at all with a four-component mixture data. Diagonal mixture models are fitted. In \fig{Random} the classes of $\bu$ are in red and in black.
As is apparent from \fig{Random}, SICL does not change the solution $K=4$ provided by BIC and ICL.

\begin{figure}[htbp]
	\begin{minipage}[c]{0.48\linewidth}
		\begin{center}
			 \includegraphics[trim=0.65cm 0.7cm 0 0, clip]{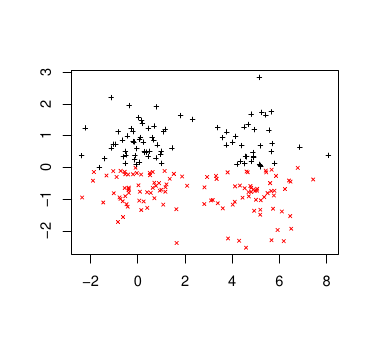}
		\end{center}
	\end{minipage}
	\begin{minipage}[c]{0.48\linewidth} 
		\begin{center}
			\includegraphics[trim=0cm 0.7cm 0 0, clip]{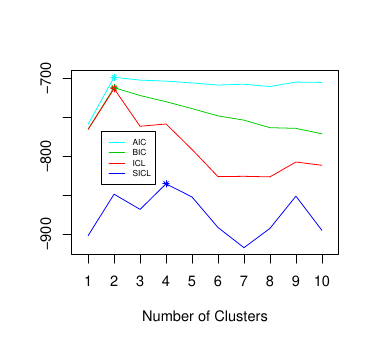}
		\end{center}
	\end{minipage}
	\caption{``Conditionally dependent labels'' data set external classification (left) and corresponding information criteria versus the number of clusters (right)}\label{NonIndep}
\end{figure}

\begin{figure}[htbp]
	\begin{minipage}[c]{0.48\linewidth}
		\begin{center}
			 \includegraphics[width = \linewidth, trim=0.65cm 1.5cm 0 0, clip]{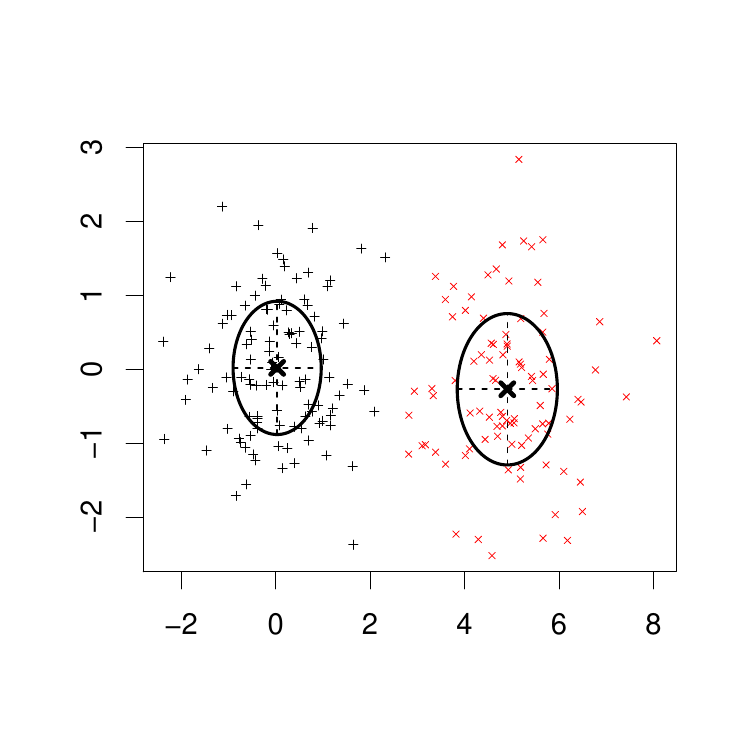}
		\end{center}
	\end{minipage}
	\begin{minipage}[c]{0.48\linewidth} 
		\begin{center}
			\includegraphics[width = \linewidth, trim=0.65cm 1.5cm 0 0, clip]{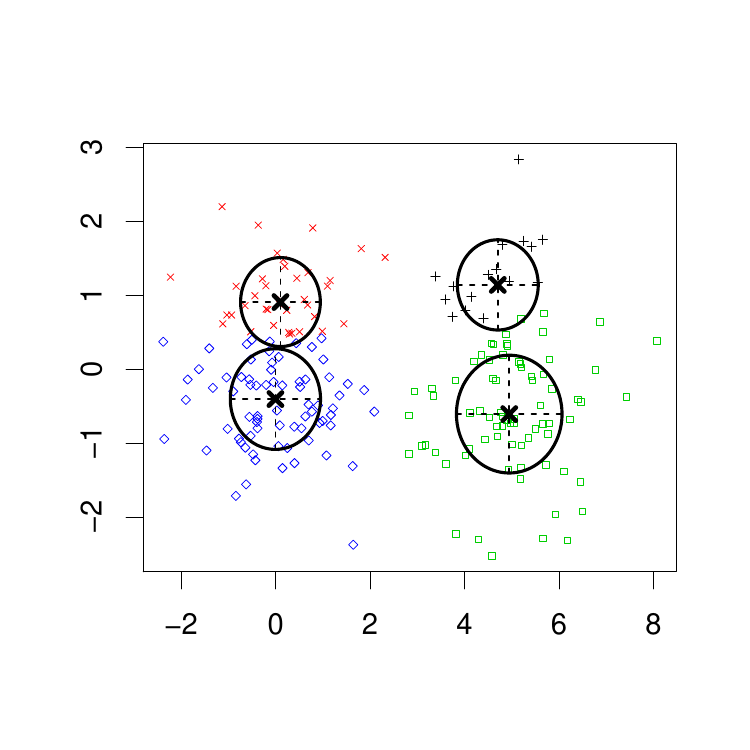}
		\end{center}
	\end{minipage}
	\caption{Two- (left) and four-cluster (right) solutions for the ``conditionally dependent labels'' data set}\label{NonIndep_24}
\end{figure}

\paragraph{About the conditional independence assumption}\label{CI} The heuristics leading to SICL assumes that $\bu$ and $\by$ are independent conditionally on $\bz$ (see Section~\ref{sic.sic}). This assumption is questionable and might not hold for all the considered models. The next experiment aims at studying the behaviour of SICL when it can be regarded as inappropriate. We consider a two-component diagonal mixture and a two-class $\bu$ partition ``orthogonal'' to this mixture. In \fig{NonIndep} the classes of $\bu$ are in red and in black. Diagonal mixture models with free volumes and proportions but fixed shapes and orientations are fitted \citep[see][]{CelGov95}. As is apparent from \fig{NonIndep} and Table~\ref{NonIndepTable}, SICL highlights the two- and the four-cluster solutions. Actually the conditional independence does not hold for the two cluster solution but it does for the four cluster solution (see \fig{NonIndep_24}) which is of interest when the focus is on the link between $\bz$ and $\bu$. However it is clear from the dispersion of the numbers of clusters selected with SICL (Table~\ref{NonIndepTable}) that this criterion is jeopardized when the conditional independence does not hold for the relevant numbers of clusters. 


\begin{table}[htbp]
	\begin{center}
		\caption{\label{NonIndepTable} Number of components selected by each criterion for the ``Conditionally dependent labels'' data.}
		\begin{tabular}{| l| c| c| c| c| c| c| c| c| c| c|}
			\hline
			K & 1 & 2 & 3 & 4 & 5 & 6 & 7 & 8 & 9 & 10 \\ 
			\hline
			AIC &0 &47 &24 &14 &4 &4 &2 &0 &4 & 1 \\
			BIC & 0 &99 &1 &0 &0 &0 &0 &0 &0 & 0 \\
			ICL &  0 &100& 0 &0 &0& 0& 0 &0 &0 & 0 \\
			SICL & 0 &36 &10 &19 &7 &6&5& 6& 7&  4 \\
			\hline
		\end{tabular}
	\end{center}
\end{table}

\subsection{Real data set: wholesale customers}\label{sim.sim.rea}
The segmentation of customers of a wholesale distributor is performed to illustrate the performance of the SICL criterion. The data set refers to 440 customers of a wholesale: $298$ from the Horeca (Hotel/Restaurant/Caf\'e) channel and $142$ from the Retail channel. They are distributed into two large Portuguese city regions (Lisbon and Oporto) and a complementary region.

\begin{table}[htbp]
	\begin{center}
		\caption{Distribution of the Region variable}\label{}
		\begin{tabular}{|c|c|c|}
		 	\hline
			{Region} & {Frequency} & {Percentage} \\
			{Lisbon} & 77 & 17.5 \\
			{Oporto} & 47 & 10.5 \\
			{Other region} & 316 & 31.8 \\
			{Total} & 440 & 100 \\
			\hline
		\end{tabular}
	\end{center}
\end{table}

The wholesale data concern customers. They include the annual spending in monetary units (m.u.) on product categories: fresh products, milk products, grocery, frozen products, detergents and paper products, and delicatessen. These variables are summarized  in \tab{Sales}.

\begin{table}[htbp]
	\begin{center}
		\caption{Product categories sales (m.u.).}\label{Sales} 
    		\begin{tabular}{|c|c|c|}
		      \hline
	      	      & Mean & Std. Deviation \\
		      Fresh products & 12000 & 12647 \\
      			Milk products & 5796 & 5796 \\
	      		Grocery & 7951 & 9503 \\
	      		Frozen & 3072 & 4855 \\
      			Detergents and Paper & 2881 & 4768 \\
	     		Delicatessen & 1525 & 2820 \\
     		 	\hline
	    \end{tabular}
    \end{center}
\end{table}

\begin{figure}[htbp]
	\begin{center}
	 	\includegraphics[width = 0.6\linewidth, trim = 0.3cm 0 0 1cm, clip]{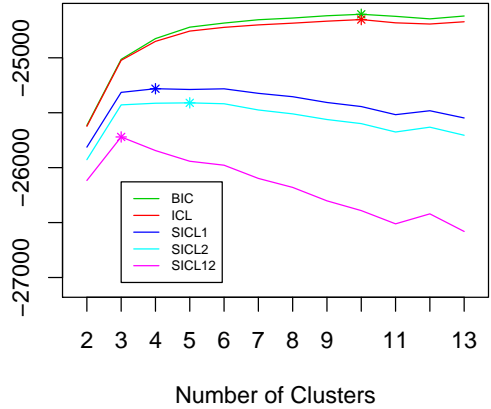}
	\end{center}
	\caption{Information criteria vs the number of clusters for the wholesale dataset}\label{WS_Criteria}
\end{figure}

These data also include responses to a questionnaire intended to evaluate possible managerial actions with potential impact on sales such as improving the store layout, offering discount tickets or extending products' assortment. The customers were asked whether the referred action would have impact on their purchases in the wholesale and their answers were registered in the scale: 1-\emph{Certainly no}; 2-\emph{Probably no}; 3-\emph{Probably yes}; 4-\emph{Certainly yes}. Diagonal mixture models have been fitted on the continuous variables described in \tab{Sales}. The results are presented in \fig{WS_Criteria}.

The SICL values based on the Channel, Region, Channel and Region external variables are indicated by SICL1, SICL2 and SICL12 respectively. BIC and ICL select a useless nine-cluster solution, with no clear interpretation. SICL1 selects a four-cluster solution, SICL2 a five-cluster solution and SICL12 a three-cluster solution.

\begin{figure}[htbp]
	\begin{minipage}[c]{0.48\linewidth} 
		\begin{center}
			\includegraphics[width = \linewidth]{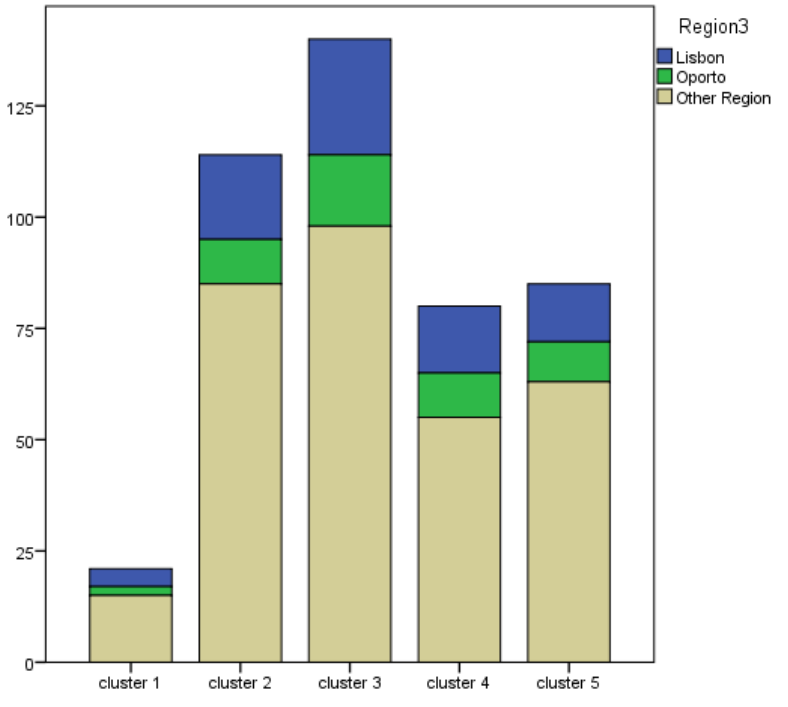}
		\end{center}
		\caption{Distribution of the variable Region on the SICL2 solution}\label{WS_5Clusters}
		\end{minipage}
	\begin{minipage}[c]{0.09\linewidth}
	\ 
	\end{minipage}
	\begin{minipage}[c]{0.48\linewidth} 
		\begin{center}
			\includegraphics[width = 1.1\linewidth]{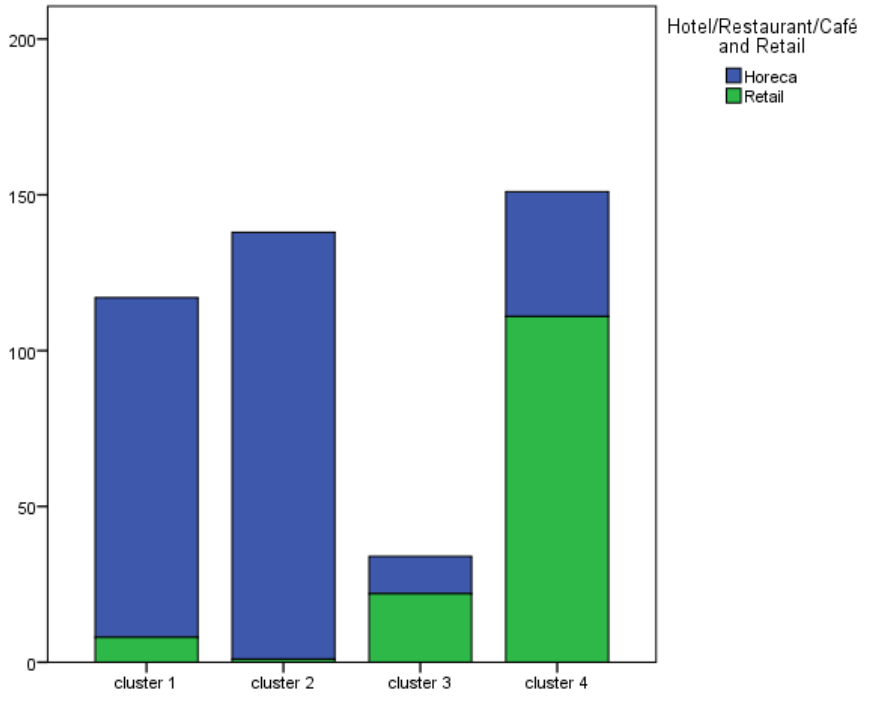}
		\end{center}
		\caption{Distribution of the variable Channel on the SICL1 solution}\label{WS_4Clusters_Channel}
	\end{minipage}
\end{figure}

\begin{figure}[htbp]
	\vspace{-1em}
	\begin{minipage}[c]{0.48\linewidth} 
		\begin{center}
			\includegraphics[width = 1.1\linewidth]{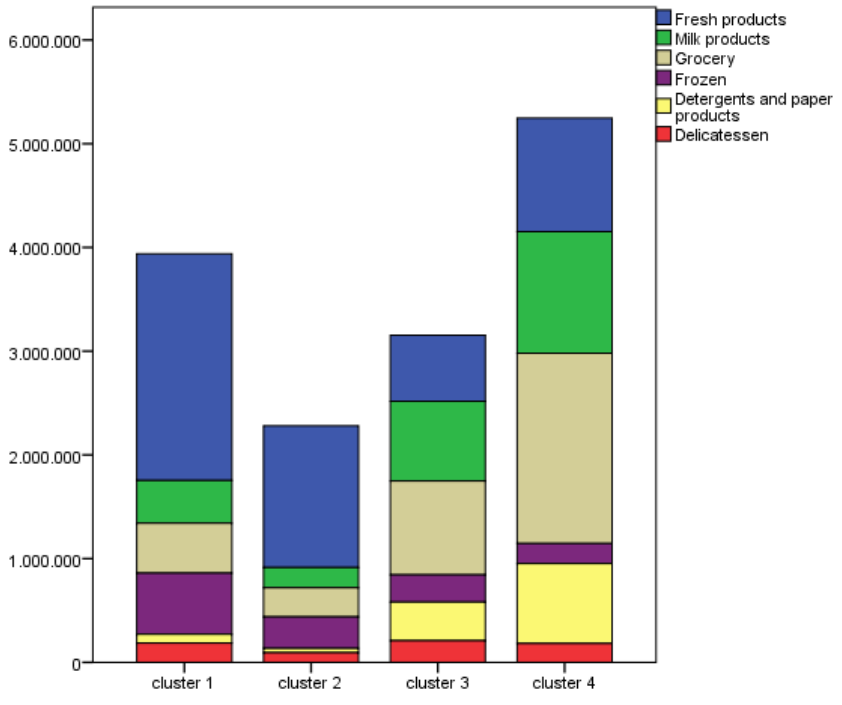}
		\end{center}
		\caption{Distribution of the product categories on the SICL1 solution}\label{WS_4Clusters}
	\end{minipage}
	\begin{minipage}[c]{0.09\linewidth}
	\ 
	\end{minipage}
	\begin{minipage}[c]{0.48\linewidth} 
		\vspace{1em}
		\begin{center}
			\includegraphics[width = 1.1\linewidth]{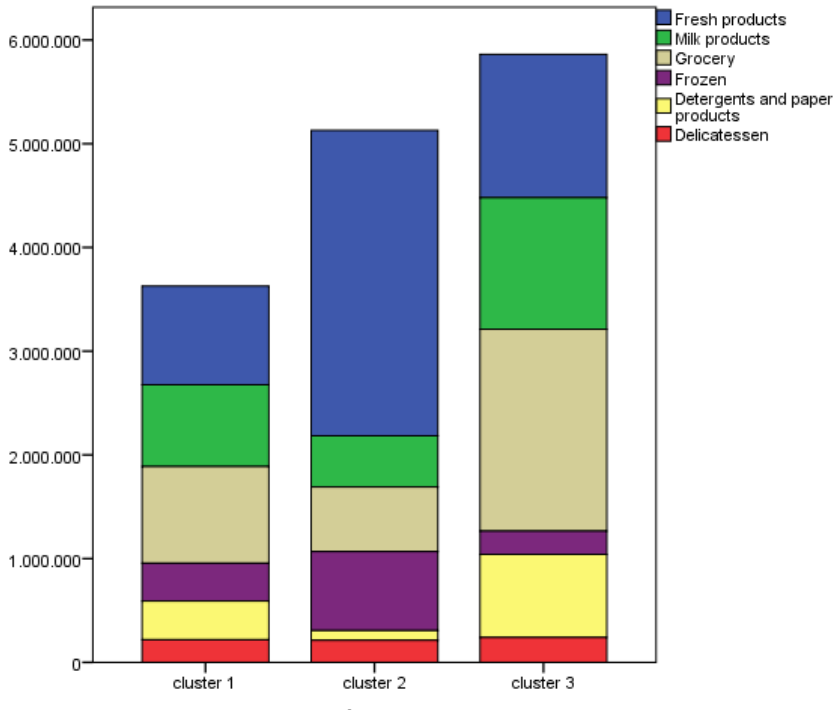}
		\end{center}
		\caption{Distribution of the product categories on the SICL12 solution}\label{WS_3Clusters}
	\end{minipage}
\end{figure}

\begin{figure}[htbp]
	\begin{center}
		\includegraphics[width = 0.6\linewidth]{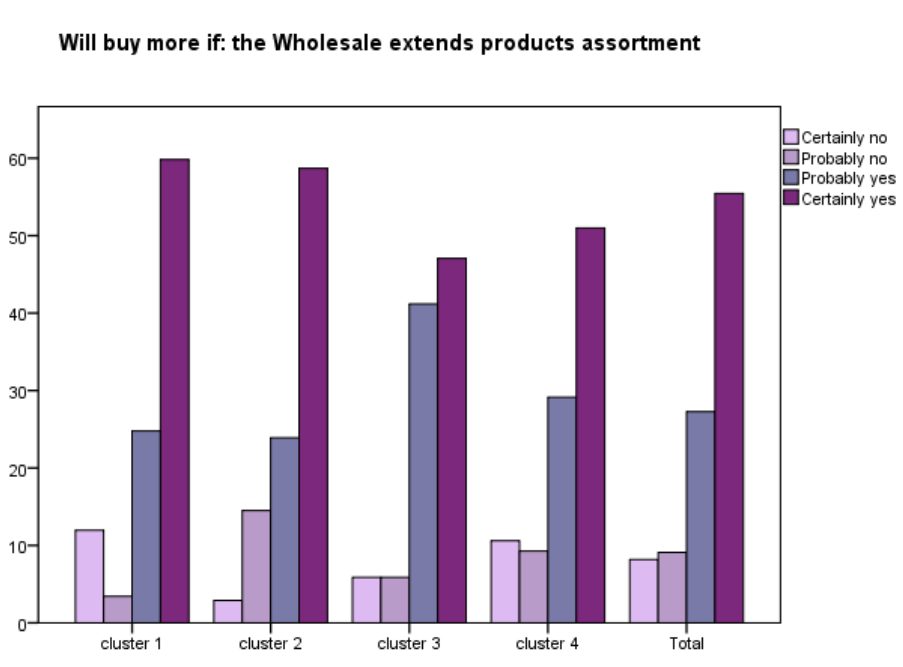}
	\end{center}
	\caption{SICL1 solution and managerial actions}\label{WS_Extend4}
\end{figure}

The five-cluster solution is less usable than the alternatives (see \fig{WS_5Clusters}). \fig{WS_4Clusters_Channel} highlights the link between the four-cluster solution and the Channel external variable. The product categories spending patterns (total spending) associated to each cluster are displayed in \fig{WS_4Clusters}. The cluster 3 is small but includes customers that spend a lot and tend to be particularly sensitive to the potential extension of the products' assortment (see \fig{WS_Extend4}).

\begin{figure}[htbp]
	\begin{minipage}[c]{0.48\linewidth} 
		\begin{center}
			\includegraphics[width = 1.1\linewidth]{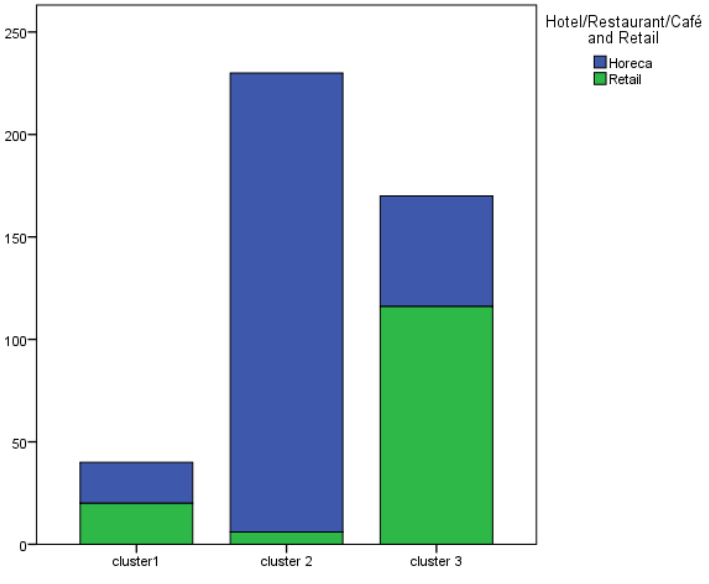}
		\end{center}
		\caption{Distribution of the Channel variable on the SICL12 solution}\label{WS_3Clusters_Channel}
	\end{minipage}
	\begin{minipage}[c]{0.09\linewidth}
  		\ 
	\end{minipage}
	\begin{minipage}[c]{0.48\linewidth}
		\begin{center}
			 \includegraphics[width = \linewidth]{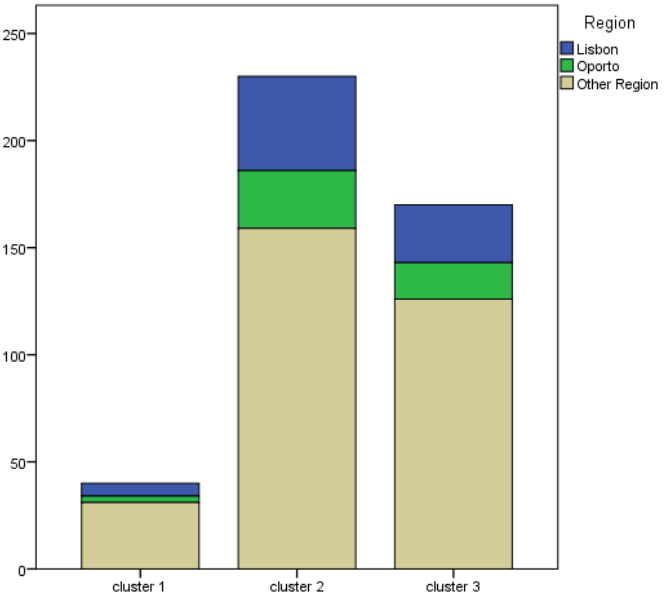}
		\end{center}
		\caption{Distribution of the Region variable on the SICL12 solution}\label{WS_3Clusters_Region}
	\end{minipage}
\end{figure}

\begin{figure}[htbp]
	\begin{minipage}[c]{0.48\linewidth}
		\begin{center}
			\includegraphics[height = 13.5em, width = 1.1\linewidth]{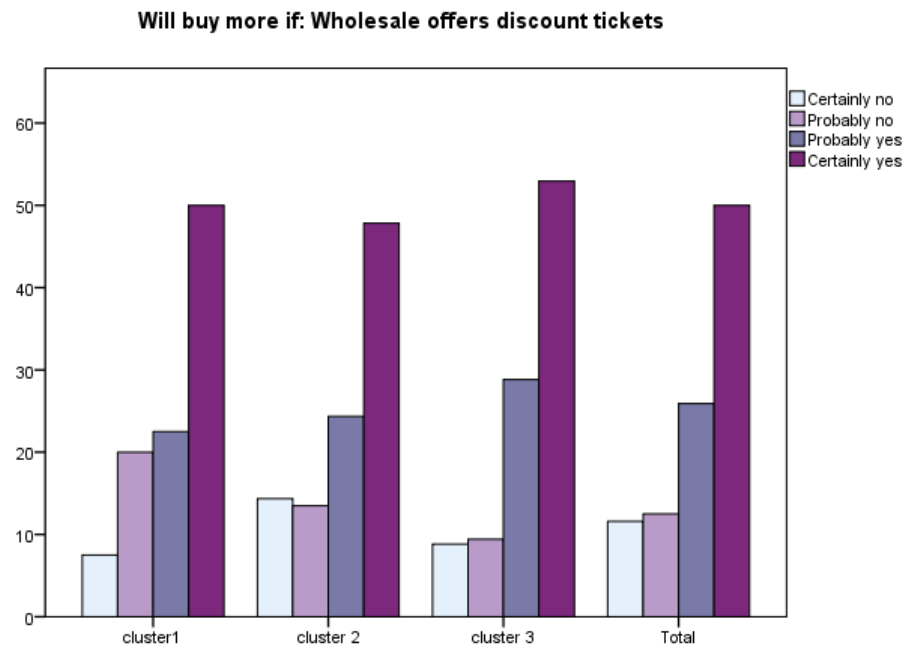}
		\end{center}
	\end{minipage}
	\begin{minipage}[c]{0.09\linewidth}
  		\ 
	\end{minipage}
	\begin{minipage}[c]{0.48\linewidth}
		\vspace{-0.5em}
		\begin{center}
			 \includegraphics[height = 14.2em, width = 1.1\linewidth]{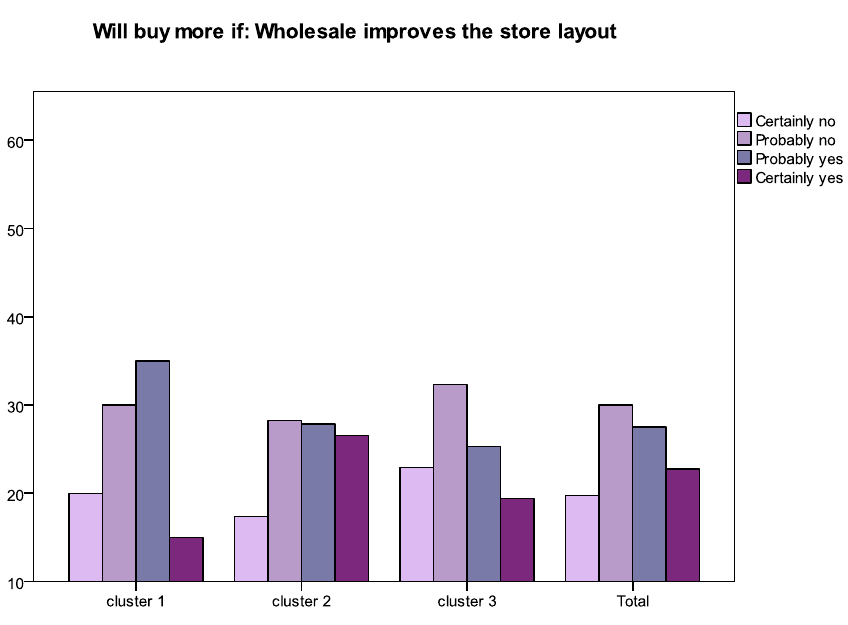}
		\end{center}	
	\end{minipage}
	\caption{SICL12 solution and managerial actions}\label{WS_SICL12_Actions}
\end{figure}

SICL12 provides the most clear-cut selection (see \fig{WS_Criteria}) and parsimonious solution. In fact, this three-cluster solution is well linked with the external variables (see Figures~\ref{WS_3Clusters_Channel} and \ref{WS_3Clusters_Region}) while the clusters remain easily discriminated by the product categories' spendings: in particular, cluster 2 (resp. 3) includes a majority of Horeca (resp. Retail) customers buying a lot of fresh products (resp. grocery) (see \fig{WS_3Clusters}). Cluster 3 is slightly more sensitive to the offering of discount tickets while cluster 2 is slightly more prone to react to improvement of the store layout (see \fig{WS_SICL12_Actions}).

\section{Discussion}\label{sic.dis}

The criterion SICL has been conceived in the model-based clustering context to
choose a sensible number of classes, possibly taking advantage of an external categorical variable or a set of external categorical variables of interest (variables other than the variables on which the clustering is based).
This criterion can be useful to draw attention to a well-grounded classification related to this
external categorical variables.

In mixture analysis several authors propose to make the mixing proportions depend on covariates through logistic models \citep[see for example][]{DayMac88}. However, to the best of our knowledge, there were no works relying on the point of view of SICL in model selection: to select the model with the help of external variables which are not used to fit each model. 

As one of the referees noticed, it may be possible to get the same solution as SICL by including the illustrative variables in the model, as it is the case for example for the Iris data set. But not including them in the model provides a stronger evidence of the link between the illustrative variables and the clustering for the very reason that the illustrative variables are not involved in the design of the clustering.

A possible limitation of SICL is the conditional independence assumption on which its heuristics rely (see Section~\ref{sic.sic}). The simulation study described at the end of Section~\ref{sic.sim.sim} illustrates the situation where $\bu$ is not conditionally independent on $\by$ given $\bz$ for the number of clusters selected by BIC or ICL. In such a case SICL could highlight clusterings for which $\bu$ and $\by$ are conditionally independent and tends to select a model with a higher number of clusters than BIC and ICL. Indeed SICL tends to select a model in which $\bz$ summarizes the information brought both by $\bu$ and $\by$. It could be expected to be far from the clusterings provided by standard model selection criteria but in doing so SICL points out a sensible and interesting solution.

Section~\ref{sim.sim.rea} and experiments not reported here suggest that the higher the number of external variables the greater the interest of SICL. 

Finally SICL could highlight partitions of special interest with respect to external categorical variables. Therefore, we think that SICL deserves to enter in the toolkit of model selection criteria for clustering. In most cases, it will propose a sensible solution and when it points out an original solution, it could be of great interest for practical purposes.


\newpage

\nocite{BieCelGovLan06}
\bibliographystyle{apalike} 
\bibliography{/Users/JP/Maths/jp.bib} 

\begin{thebibliography}{}

\bibitem[Baudry, 2009]{Bau09}
Baudry, J.-P. (2009).
\newblock {\em Model Selection for Clustering. Choosing the Number of Classes.}
\newblock PhD thesis, Univ. Paris-Sud.
\newblock \url{http://tel.archives-ouvertes.fr/tel-00461550/fr/}.

\bibitem[Biernacki et~al., 2000]{BieCelGov00}
Biernacki, C., Celeux, G., and Govaert, G. (2000).
\newblock Assessing a mixture model for clustering with the integrated
  completed likelihood.
\newblock {\em IEEE Trans. PAMI}, 22:719--725.

\bibitem[Biernacki et~al., 2006]{BieCelGovLan06}
Biernacki, C., Celeux, G., Govaert, G., and Langrognet, F. (2006).
\newblock {Model-based cluster and discriminant analysis with the mixmod
  software}.
\newblock {\em Computational Statistics and Data Analysis}, 51(2):587--600.

\bibitem[Celeux and Govaert, 1995]{CelGov95}
Celeux, G. and Govaert, G. (1995).
\newblock {G}aussian parsimonious clustering models.
\newblock {\em Pattern Recognition}, 28(5):781 -- 793.

\bibitem[Dayton and Macready, 1988]{DayMac88}
Dayton, C.~M. and Macready, G.~B. (1988).
\newblock Concomitant-variable latent-class models.
\newblock {\em Journal of the American Statistical Association},
  83(401):173--178.

\bibitem[Dempster et~al., 1977]{DemLaiRub77}
Dempster, A., Laird, N., and Rubin, D. (1977).
\newblock Maximum likelihood from incomplete data via the {EM}-algorithm.
\newblock {\em Journal of the Royal Statistical Society. Series B},
  39(1):1--38.

\bibitem[Fisher, 1936]{Fis36}
Fisher, R. (1936).
\newblock The use of multiple measurements in taxonomic problems.
\newblock {\em Annals of Eugenics}, 7:179--188.

\bibitem[Keribin, 2000]{Ker00}
Keribin (2000).
\newblock Consistent estimation of the order of mixture models.
\newblock {\em Sankhya A}, 62(1):49--66.

\bibitem[Lebret et~al., 2012]{LebIovLanBieCelGov12}
Lebret, R., Iovleff, S., Langrognet, F., Biernacki, C., Celeux, G., and
  Govaert, G. (2012).
\newblock Rmixmod: The r package of the model-based unsupervised, supervised
  and semi-supervised classification mixmod library.
\newblock \url{http://cran.r-project.org/web/packages/Rmixmod/index.html}.

\bibitem[McLachlan and Krishnan, 1997]{MclKri97}
McLachlan, G. and Krishnan, T. (1997).
\newblock {\em The {EM}-algorithm and Extensions}.
\newblock New York : Wiley.

\bibitem[McLachlan and Peel, 2000]{MclPee00}
McLachlan, G. and Peel, D. (2000).
\newblock {\em Finite Mixture Models}.
\newblock New York : Wiley.

\bibitem[Roeder and Wasserman, 1997]{RoeWas97}
Roeder, K. and Wasserman, L. (1997).
\newblock Practical bayesian density estimation using mixtures of normals.
\newblock {\em Journal of the American Statistical Association},
  92(439):894--902.

\bibitem[Schwarz, 1978]{Sch78}
Schwarz, G. (1978).
\newblock Estimating the dimension of a model.
\newblock {\em Ann. Statist.}, 6:461--464.

\end{thebibliography}

\newpage

\

\vfill

\noindent \emph{Jean-Patrick Baudry}
\\Universit\'e Pierre et Marie Curie - Paris VI \\
Bo\^ite 158, Tour 15-25, 2\textsuperscript{e} \'etage \\
4 place Jussieu, 75252 Paris Cedex 05\\ 
France.\\ 
Jean-Patrick.Baudry@upmc.fr\\
\url{http://www.lsta.upmc.fr/Baudry}

\end{document}